\documentclass{epl}
\usepackage{amsmath}
\usepackage{amssymb}
\usepackage{latexsym}
\newcommand{\radius}{a}
\newcommand{\bk}{{\bf k}}

\newcommand{\e}{\mbox{\large e}}
\newcommand{\dd}{{\mbox{d}}}
\newcommand{\kb}{k_{\mbox{\scriptsize B}}}
\newcommand{\diff}[2]{\frac{\mbox{d} #1}{\mbox{d} #2 } }

\title{Nonlinear viscoelasticity of metastable complex fluids}
\author{Kunimasa Miyazaki\inst{1}
\and
Hans M. Wyss\inst{2}
\and
David A. Weitz\inst{2}
\and
David R. Reichman\inst{1}
}
\institute{
\inst{1} Department of Chemistry,
Columbia University, 3000 Broadway, New York, NY 10027\\
\inst{2} Department of Physics \& DEAS, Harvard University,
Cambridge, MA 02138
}


\begin{document}

\maketitle

\begin{abstract}
 Many metastable complex fluids such as colloidal glasses and gels
 show distinct nonlinear viscoelasticity with increasing
 oscillatory-strain amplitude; the storage modulus
 decreases monotonically as the strain amplitude increases
 whereas the loss modulus has a distinct peak before it decreases
 at larger strains.
 We present a qualitative argument to explain this ubiquitous behavior
 and use mode coupling theory (MCT) to confirm it.
 We compare theoretical predictions to the measured
 nonlinear viscoelasticity in a
 dense hard sphere colloidal suspensions;
 reasonable agreement is obtained.
 The argument given here can be used to obtain new information
 about linear viscoelasticity of metastable complex fluids
 from nonlinear strain measurements.
\end{abstract}

Rheological measurements are commonly used to investigate dynamical
and structural properties of complex fluids. Oscillatory linear
viscoelastic experiments measure the response function, or the
complex modulus,
$G^{\ast}(\omega)=G'(\omega)+iG''(\omega)$~\cite{larson1999}.
This provides information about the relaxation of the microstructure of
the system as a function of oscillation frequency, $\omega$. This
function is strictly defined only within the linear response regime,
and thus an important practical experimental issue is the determination of its
range of validity. However, many complex fluids exhibit a nonlinear
response even at very small applied strains.
This can provide an alternate means of probing the properties of the
system; the strain amplitude, $|\gamma|$, can be used as an independent
parameter~\cite{Macosko1993,Hyun2002}.
While the nonlinear rheological properties of complex fluids are very
diverse~\cite{larson1999}, most colloidal suspensions and polymer
solutions show a simple yielding behavior, where both the storage
($G'$) and loss ($G''$) modulus decrease monotonically as $|\gamma|$
increases. Less common is strain hardening (or strain-stiffening),
where $G'$ increases with $|\gamma|$. This can be observed in
polymer solutions with strongly cross-linked
chains~\cite{larson1999}, soft-biological
tissues~\cite{Gardel2004,Storm2005}, and in dense colloidal
suspensions~\cite{Bertrand2002,Melrose2004}.
There is another type of observed rheological
behavior that lies between these two kinds of responses: as $|\gamma|$
increases, $G'$ continuously decreases above a critical strain, whereas
$G''$ exhibits a distinct peak before decreasing at even larger applied
strains. This behavior is
ubiquitous in metastable complex fluids. It was first reported in
carbon black solutions~\cite{Payne1963}; many other systems exhibit
similar behavior, including polymer
systems~\cite{Tirtaatmadja1997a}, fumed
silica~\cite{Raghavan1995,Yziquel1999}, colloidal
glasses~\cite{Mason1995,Petekidis2002},
emulsions~\cite{Mason1995b,Bower1999}, gels~\cite{Altmann2004} and
electrorheological fluids~\cite{Parthasarathy1999,Sim2003}. Despite
the diversity of these systems, the observed behavior is strikingly
universal: (i) The peak in $G''$ appears around $10^{-2} \lesssim
|\gamma| \lesssim 1$ over a wide range of frequencies; (ii) the peak
in $G''$ is observed for a frequency regime where the system is
elastic, with $G' > G''$; and (iii) the shear thinning exponent
$\nu$ of $G'' \propto |\gamma|^{-\nu}$ at larger strains is always
about one half of the exponent associated with $G'$. There have been
many attempts to explain these universally observed phenomena. It
has been argued that the increase of $G''$ as a function of
$|\gamma|$ is due to the increase of the effective volume of the
temporal network structure in polymer
solutions~\cite{Tirtaatmadja1997a}. Alternative suggestions include
the breakdown of aggregates into small clusters which are more
dissipative~\cite{Yziquel1999} and the rearrangement of clusters due
to the strain associated with thickening~\cite{Parthasarathy1999}.
The phenomenological soft glassy rheology (SGR) model also accounts
for the peak in $G''$ as a function of $|\gamma|$, and posits that
flow near the glass transition arises from activated barrier
crossing~\cite{Sollich1998}. However, a full microscopic account of this
ubiquitous behavior is still lacking.

In this Letter, we show that the origin of the $|\gamma|$-dependent
behavior of $G^{\ast}$ is in fact remarkably general and transcends
the physical ingredients of the SGR picture;
a peak in $G''$ should be present in any viscoelastic complex fluid
that behaves like an elastic solid at small strain but yields and flows at
larger strains.
In particular, we explicitly demonstrate that this is true even in the
absence of local activated processes.
To make this
argument more quantitative, we focus on colloidal suspensions in the
supercooled state and calculate the $\omega$ and $|\gamma|$
dependence of $G^{\ast}$ using mode-coupling theory (MCT),
generalized to account for the presence of oscillatory shear strain.
This method successfully accounts for the peak observed in our
experiments on a hard sphere colloidal suspension near its glass
transition density.

The presence of the peak can be qualitatively explained using the
following simple argument.
Throughout the paper, an oscillatory strain,
$\gamma(t) =|\gamma|\e^{i\omega t}$, is considered.
In the small strain limit, the stress is written as
$\sigma(t)= G^{\ast}(\omega)\gamma(t)$.
As the simplest example, we consider a Maxwell model,
where the storage and loss moduli are given by
\begin{equation}
G'(\omega) \propto \frac{(\omega\tau_0)^2}{1 + (\omega\tau_0)^2}
\hspace*{0.5cm}\mbox{and}\hspace*{0.5cm}
G''(\omega) \propto \frac{\omega\tau_0}{1 + (\omega\tau_0)^2},
\label{eq:maxwell}
\end{equation}
respectively.
Here $\tau_0$ is a characteristic relaxation time
which in many systems represents the structural relaxation time.
According to this equation, $G''(\omega)$ has a peak around
$\omega \simeq 1/\tau_0$.
By contrast, for large strains many complex fluids experience a drastic
decrease of their structural
relaxation time.
Such a decrease can be characterized by the strain rate amplitude,
$|\dot{\gamma}|$, and is given by
the phenomenological expression
\begin{equation}
1/\tau \simeq 1/\tau_0 + K|\dot{\gamma}|^{\nu},
\label{eq:tau}
\end{equation}
where $\nu$ is a positive exponent and $K$ is a constant. If
$\tau_0$ is replaced by $\tau$ in Eq.~(\ref{eq:maxwell}), a peak of
$G''$ should be observed  at
$\omega \simeq 1/\tau \simeq K|\dot{\gamma}|^{\nu}$;
this equality is satisfied if $\omega \gg 1/\tau_{0}$
and the system's intrinsic relaxation is very slow (thus
$1/\tau_0$ can be neglected). In particular, metastable systems
which show slow dynamics, such as supercooled fluids and gels, have
an exponent $\nu$ very close to unity and $K$ of order
1~\cite{miyazaki2002,miyazaki2004}; then, the peak always appears
near $\omega \simeq |\dot{\gamma}|$ or $|\gamma| \simeq 1$. This
argument also accounts for the facts that (i) the position of the
peak is relatively insensitive to the frequency, (ii) the peak of $G''$ should
appear at frequencies where the linear viscoelasticity is
dominantly elastic, ($G'> G''$), and (iii) the shear thinning
exponents at larger strains are related and are given by $G' \simeq
|\gamma|^{-2\nu}$ and $G'' \simeq |\gamma|^{-\nu}$. Considering the
broad diversity of systems, it is remarkable that this simple
argument correctly captures the features of the peak in $G''$.
The above argument is based on the assumption that nonlinear effects
solely stem from the strain amplitude $|\gamma|$ (or more precisely
from the strain rate amplitude $|\dot{\gamma}|=\omega|\gamma|$) and are
independent of the time-dependence of the strain.
In other words, the stress is given by $\sigma(t)=G^{\ast}(\omega,
|\gamma|)\gamma(t)$.
This is consistent with recent experiments which have taken the effect
of higher order harmonics into account and shown
that they account for at most 10\% of the
response~\cite{Wilhelm2002,Hyun2002}, thus, lending support to the
notion these terms are
negligible.

In order to make this argument more quantitative,
we use  a
microscopic description based on MCT. First, we focus on a typical
complex fluid which shows a peak in $G''$ as a function of
$|\gamma|$, namely, a colloidal suspension in the supercooled state.
To date, MCT is the only first principles theory that describes the
dynamics of the supercooled state. Recently, several attempts have
been made to
extend the MCT approach to systems under shear strain
for either stationary
systems~\cite{miyazaki2002,miyazaki2004,Szamel2004} or for transient
states~\cite{fuchs2002,fuchs2005}. In all previous analyses,
$\dot{\gamma}$ has been assumed to be time-independent. Here we
generalize the analysis~\cite{miyazaki2002,miyazaki2004} to
time-dependent oscillatory strain.
We consider strain in the
$x$ direction given by $\gamma(t) =|\gamma|\sin(\omega t)$.
The arguments used are exactly the same as for the constant shear
case\cite{miyazaki2002,miyazaki2004},  which should be valid
as long as the shear rate amplitude $\omega|\gamma|$ is small compared to the
inverse of the diffusion time $a^2/D$.
We define a density correlation function which satisfies generalized
translational invariance, $F(k,t) = N^{-1} \langle \delta\rho_{{\bf
k}(-t)}(t)\delta\rho_{{\bf -k}}(0) \rangle$, where $N$ is the total
particle number and $\bk(t)$ is the time dependent wavevector
defined as $\bk(t) = (k_x, k_y+\gamma(t)k_x,
k_z)$~\cite{miyazaki2002}. Then, the MCT equation for $F(k,t)$ is
\begin{equation}
\begin{aligned}
\diff{ F(\bk, t) }{t}
=
&
-\frac{Dk(-t)^2}{S(|\bk(-t)|)}F(\bk, t)
-\int_{0}^{t}\!\!\dd t'~
  M(\bk(-t),t-t')
  \diff{ F(\bk, t') }{t'},
\end{aligned}
\label{eq:mct1}
\end{equation}
where $D$ is the diffusion constant in the dilute limit, $k =
|\bk|$, and $S(k)$ is the (unperturbed) static structure factor.
$M(\bk,t)$ is the MCT memory kernel and is a nonlinear function of
$F(k,t)$ and $S(k)$, whose explicit expression is given in
Refs.~\cite{miyazaki2002,miyazaki2004}. Solving Eq.~(\ref{eq:mct1})
numerically is a demanding task because of the anisotropic nature of
the equation in the presence of the unidirectional strain. In order
to make the computation tractable, we adopt the ``isotropically
sheared hard sphere model'' (ISHSM) approximation proposed by Fuchs
{\it et al.}~\cite{fuchs2002}, where the anisotropy of the strain
which appears in ${\bf k}(t)$ is neglected and the scalar $|{\bf
k}(t)|$ is approximated by $|\bk(t)| = \sqrt{k^2 + k_xk_y \gamma(t)
+ k_y^2 \gamma^2(t)} \simeq \sqrt{1 + {\gamma^2(t)}/{3}}~k$. This
approximation is expected to be qualitatively correct as verified by
the observation that the dynamics of supercooled liquids under shear
strain is surprisingly isotropic~\cite{miyazaki2004}. We solve
Eq.~(\ref{eq:mct1}) using this approximation and evaluate $F(k,t)$
for a given set of $\omega$ and $|\gamma|$. The complex modulus is
calculated in the same manner as $M(k, t)$ in eq.(\ref{eq:mct1}) and
is given by\cite{miyazaki2004}
\begin{equation}
G_{\mbox{\scriptsize NL}}^{\ast}(\omega, |\gamma|) =
\frac{\kb T i\omega}{2}\int_{0}^{\infty}\!\!\dd t
\int\!\!\frac{\dd \bk}{(2\pi)^3}~\e^{i\omega t}V(k)F^2(k,t)V(|\bk(t)|),
\end{equation}
where $V(k)= k_x S^{-2}(k)nS^{\prime}(k)$ and $n$ is the particle number
density.
We calculate
$G_{\mbox{\scriptsize NL}}^{\ast}(\omega;\gamma)$ for a hard sphere
colloidal suspension
with particle radius $\radius$ for a volume
fraction $\phi=4n\pi\radius^3/3$ near the MCT glass transition point,
$\phi_c \approx 0.516$,
at which point $F(k,t)$ undergoes an ergodic-nonergodic
transition~\cite{gotze1992}. In Fig.~1, we show the frequency and
strain dependence of
$G_{\mbox{\scriptsize NL}}^{\ast}(\omega,|\gamma|)$.
The volume fraction is set to
$\phi=\phi_c\times(1-10^{-3})$. We calculate $S(k)$ from the
Percus-Yevick equation. The frequency is scaled by $4\radius^2/D$ and
$G_{\mbox{\scriptsize NL}}^{\ast}$ is plotted in units of $\kb
T/8\radius^3$.
\begin{figure}
\twoimages[scale=0.6]{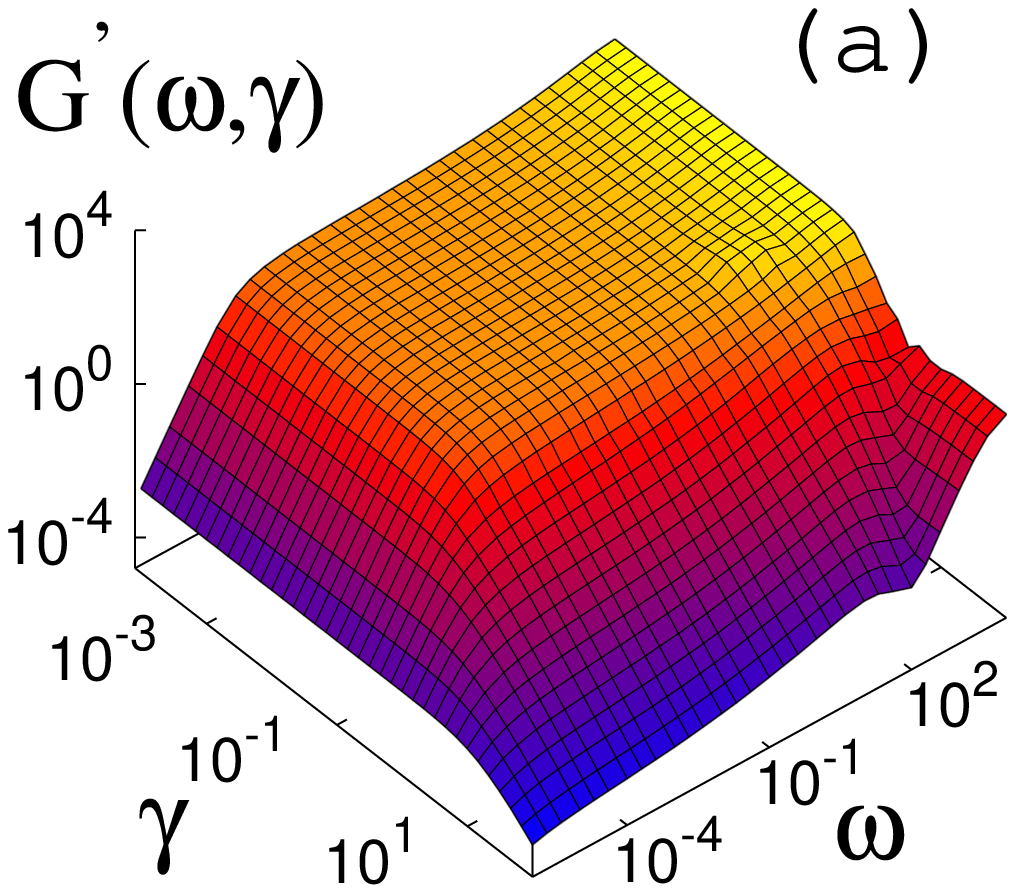}{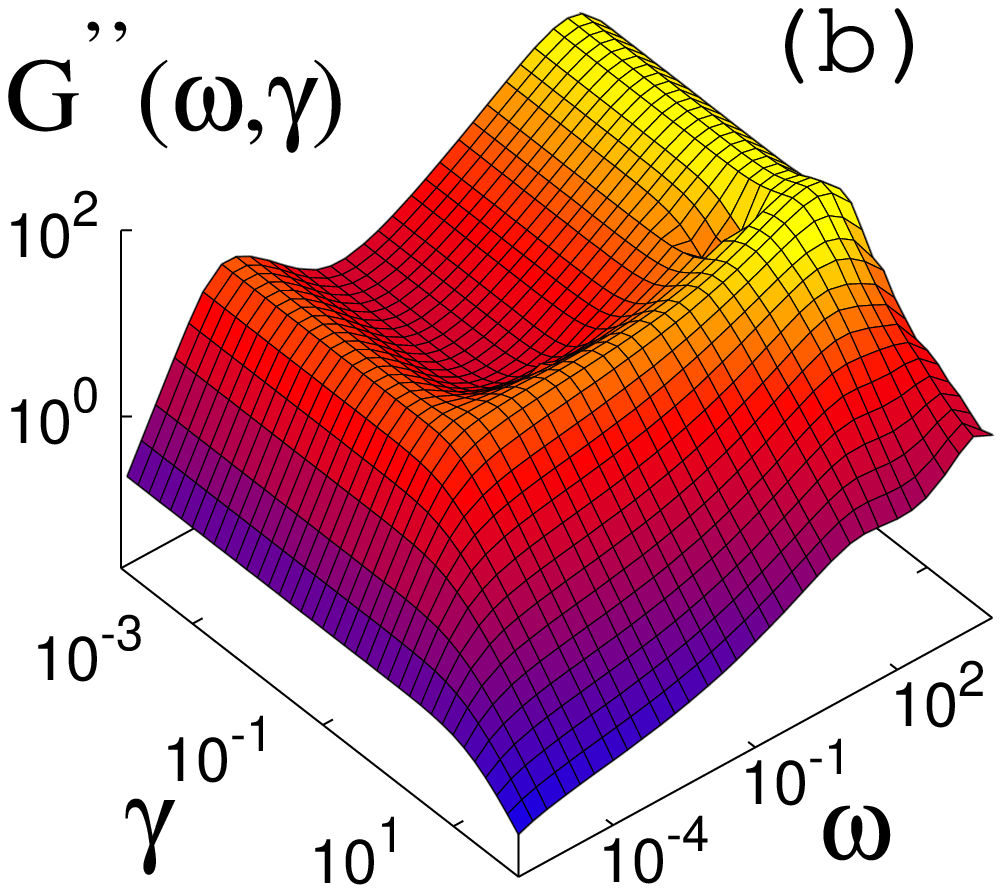}
\caption{Frequency and strain dependence of (a) the storage modulus
$G_{\mbox{\scriptsize NL}}^{\prime}(\omega;\gamma)$ and
(b) the loss modulus
$G_{\mbox{\scriptsize NL}}^{\prime\prime}(\omega;\gamma)$
 for hard sphere suspension near the glass transition density.}
\label{fig1}
\end{figure}
The behavior of $G_{\mbox{\scriptsize NL}}''$
demonstrates that, in the small strain limit, a broad dip
($\beta$-regime) is observed between $10^{-3} \lesssim \omega
\lesssim 10^{1}$. A peak at low frequencies $\omega \simeq 10^{-4}$
is the $\alpha$-relaxation regime below which liquid-like relaxation
is observed. In the $\beta$-regime,
$G_{\mbox{\scriptsize NL}}^{\prime}$ is larger than
$G_{\mbox{\scriptsize NL}}^{\prime\prime}$
and the system is solid-like. As the strain is
increased in this regime, a weak peak in
$G_{\mbox{\scriptsize NL}}^{\prime\prime}$
is observed around
$|\gamma| \simeq 0.4$,
whereas
$G_{\mbox{\scriptsize NL}}^{\prime}$
starts to decrease
monotonically. At even larger strains,
$G_{\mbox{\scriptsize NL}}^{\prime\prime}$
also starts to decrease. The decrease of
$G_{\mbox{\scriptsize NL}}^{\ast}$ is well described by a power law,
$G_{\mbox{\scriptsize NL}}'\simeq |\gamma|^{-2\nu}$
and
$G''_{\mbox{\scriptsize NL}} \simeq |\gamma|^{-\nu}$
with
$\nu=0.9$.
The features that appear at both high frequencies and strain amplitudes
occur in the regime where $\omega|\gamma|$
(or the P\'eclet number $\omega|\gamma|a^2/D$) becomes very large
and thus the validity  of the present MCT framework
is questionable.
Thus, they should not be considered as physical.
The position of the peak appears to be almost frequency independent.
Although $\nu$ varies with the system, these general features are shared
by most experiments where the peak is
observed~\cite{Payne1963,Tirtaatmadja1997a,Raghavan1995,Yziquel1999,Mason1995,Petekidis2002,Mason1995b,Bower1999,Altmann2004,Parthasarathy1999,Sim2003}.
It should be emphasized however that the peak in
$G_{\mbox{\scriptsize NL}}^{\prime\prime}$ as a function of $|\gamma|$
is not generic, intrinsic shear
thickening~\cite{Melrose2004,Storm2005} but originates instead from a
strong decrease of the structural relaxation
($\alpha$-relaxation) time at large strains. Indeed, the
$\alpha$-relaxation time, $\tau_{\alpha}$, obtained from $F(k,t)$ is
found to behave as $\tau_{\alpha}\simeq |\dot{\gamma}|^{-\nu}$.
Furthermore, our results show that the origin of the peak in
$G_{\mbox{\scriptsize NL}}^{\prime\prime}$ is more general than that
posited by the SGR model~\cite{Sollich1998}.
In particular, MCT does not capture the local rearrangements of
particles that occur via activated barrier crossing, but  it nevertheless properly captures
the features of $G_{\mbox{\scriptsize NL}}^{\prime\prime}$.
We thus conclude that activated processes are not necessary
to describe these features.

While these calculations have been done for oscillatory strain, we
note that the time-dependence of the generated strain, $\gamma(t)$,
has only a small effect on the $\omega$ and $|\gamma|$ dependence of
$G_{\mbox{\scriptsize NL}}^{\ast}$.
To show this, we calculate
$G_{\mbox{\scriptsize NL}}^{\ast}$ by approximating the oscillatory
strain rate by a constant, $\dot{\gamma}(t) \simeq \omega|\gamma|$.
$G_{\mbox{\scriptsize NL}}^{\ast}$ thus obtained is qualitatively
similar to that calculated for the case of oscillatory strain with a
slight shift to a lower strain amplitude ($|\gamma| \approx 0.3$) 
of the peak position.
These
conclusions support our qualitative argument based on the Maxwell
model. The qualitative argument also asserts that $G^{\ast}(\omega)$
is a function of $\hat{\omega}\equiv \omega\tau$ (see
Eq.~(\ref{eq:maxwell})) and the effect of the large strain comes in
only through the relaxation time in the form of Eq.~(\ref{eq:tau}).
We find that this argument is valid for structural
relaxation in glassy systems within the approximate framework
of the MCT described in this paper.
$\tau_{\alpha}(\dot{\gamma})$
extracted from the peaks of $G_{\mbox{\scriptsize
NL}}^{\prime\prime}$ shown in Fig.1 (b) as a function of the strain
rate is well described by Eq.~(\ref{eq:tau}) with
$\tau_0 =\tau_{\alpha}(\dot{\gamma}\rightarrow 0)$,
$\nu=0.9$ and a positive
constant $K$ ($\simeq2.5$ in units of $4\radius^2/D$). We also find
that, around the peaks, the shape of $G_{\mbox{\scriptsize
NL}}^{\ast}(\omega;\gamma)$ as a function of $|\dot{\gamma}|$
matches well with $G^{\ast}(\omega\tau_{\alpha})$ calculated at
$|\dot{\gamma}|=0$ as a function of $\omega\tau_\alpha$, where the
strain dependence of $\tau_{\alpha}$ is given by Eq.~(\ref{eq:tau}).

To demonstrate the power of the MCT approach, we make a quantitative
comparison of theory with an experimental system, namely, a hard
sphere colloidal suspension. Experiments are performed with
sterically stabilized PMMA particles suspended in a mixture of
decaline and cycloheptyle bromide. Their radius is $\radius$=197 nm
and their volume fraction is $\phi \approx 0.56$, which is a
super-cooled fluid, just below the glass transition. Rheological
experiments are performed in this supercooled state at room
temperature using a strain-controlled rheometer (ARES, TA
Instruments). Results for strain-sweep measurements at a fixed
frequency $\omega = 0.2$ rad/sec are shown in Fig.~2. A distinct
peak is observed in $G_{\mbox{\scriptsize NL}}^{\prime\prime}$ at
$|\gamma|\approx 0.15$, while the onset of a monotonic decrease of
$G_{\mbox{\scriptsize NL}}^{\prime}$ is observed at the same strain
amplitude. The shear thinning behavior is well described by a power
law, with a shear thinning exponent $\nu\approx 0.7$. In the linear
response regime where the strain is small, MCT is known to be
successful at predicting the frequency dependence quantitatively and
the overall shape of the complex modulus qualitatively near the
glass transition points. Due to the generically approximate nature
of MCT, a direct comparison to experimental data still requires
several fitting parameters. It is known that MCT underestimates the
glass transition volume fraction by more than
10\%\cite{vanMegen1993} (experiments predict the glass transition
density around $\phi_g\approx 0.58$, whereas MCT predicts the
nonergodic transition at $\phi_c\approx 0.516$). 
MCT also underestimates the amplitude of
the complex modulus over a wide range of frequencies, as
theoretically shown for a semi-dilute
suspension\cite{cichocki1991,verberg1997}. Moreover, MCT entirely
neglects the effects of hydrodynamic interactions. In order to
determine the parameters to correct for the above-mentioned
problems, we first perform a linear viscoelastic measurement for
wide range of frequencies ($10^{-2} \leq \omega \leq 10^2$ rad/sec)
in the small-strain limit ($|\gamma|=0.005$ and $0.01$). Then, we
regard the volume fraction as a free parameter and determine it in
such a way that the frequency dependence of $G_{\mbox{\scriptsize
NL}}^{\ast}(\omega, |\gamma|=0)$ calculated by MCT in the absence of
strain matches with the shape of the measured $G^{\ast}(\omega,
|\gamma|\ll 1)$. The volume fraction thus determined is
$\phi=\phi_c\times(1-10^{-2.3})$. The hydrodynamic interactions are
approximately taken into account by replacing the diffusion constant
$D$ in eq.(\ref{eq:mct1}) by $D_{\mbox{\scriptsize HI}}(k)$, the
short time collective diffusion coefficient with full hydrodynamic
interactions, which was calculated using the Beenakker-Mazur
theory\cite{beenakker1984}. Finally, the amplitude of $G^{\ast}$ is
matched with experimental data in the $\gamma\rightarrow 0$ limit.
For these parameters, we calculate $G_{\mbox{\scriptsize
NL}}^{\ast}$ for finite strains. The results thus obtained reproduce
the peak in $G_{\mbox{\scriptsize NL}}^{\prime\prime}$ but
overestimate the position of the peak. The lines in Fig.~2 are
obtained by using the MCT calculations with a strain that is
increased by a factor of approximately 3, so that the peak overlaps
with the experimental data. The discrepancy of the peak position
might stem largely from the ISHSM approximation which we introduced
for technical reasons.
\begin{figure}
\onefigure[scale=0.5,angle=-90]{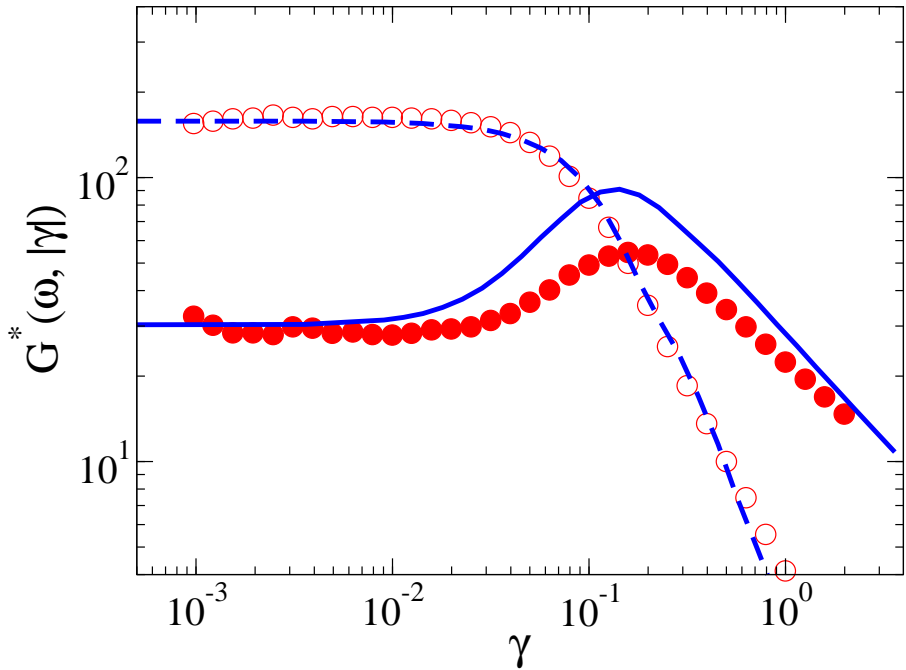}
\caption{
Comparison of MCT with experiment for a dense hard sphere suspension
for $\omega=0.2$ rad/sec.
The empty and filled circles represent the storage modulus,
$G_{\mbox{\scriptsize NL}}'$,
and loss modulus, $G_{\mbox{\scriptsize NL}}''$, respectively.
Bold lines represents the MCT results using the ISHSM  approximation
 near the glass transition density.
The lines are shifted in the direction of the strain by a factor of 3.
The modulus is plotted in units of $\kb T/8\radius^3$.
}
\label{fig2}
\end{figure}
Indeed, it is obvious that the ISHSM approximation underestimates the
onset of shear thinning because it neglects the term linear in
$|\dot{\gamma}|$.
The shear thinning after the peak exponent from MCT is found to be $\nu\simeq
0.9$ for $G_{\mbox{\scriptsize NL}}^{\prime\prime}$ and $2\nu$ for
$G_{\mbox{\scriptsize NL}}^{\prime}$.
While MCT overestimates of the shear thinning exponent and the height
and position of the peak which is somewhat larger than that extracted
from experiment, the overall shape and qualitative behavior of
$G_{\mbox{\scriptsize NL}}^{\ast}$ is captured well.
Considering the known problems associated with MCT
even in the linear regime, the result shown in Fig.~2 generated by our
``nonequilibrium'' MCT is rather remarkable.

Although in this Letter we have investigated a limited class of
systems where the MCT approach may be applicable, we believe that
our analysis is more general and gives significant insight into the
underlying physics of the peak in the loss modulus as a function of
strain, observed in various types of complex fluids. The major
origin of the peak in the loss modulus is the strong decrease of the
structural relaxation time at large strains, and has no connection
whatsoever with intrinsic shear thickening. This peak should always
be observed for systems which exhibit slow dynamics and possess a
broad band of solid-like behavior above a crossover frequency where
the storage modulus is larger than the loss modulus. The theoretical
argument predicts that, near the peak of $G''$, the strain and
frequency dependent viscoelasticity is well approximated by the
linear viscoelasticity
$G^{\ast}(\omega\tau_{\alpha}(|\dot{\gamma}|))$, as a function of
the frequency scaled by the relaxation time $\tau_\alpha$, whose
strain dependence is given by Eq~.(\ref{eq:tau}). This implies that
nonlinear measurements are a promising alternative for probing the
structure and dynamics of systems where linear viscoelastic
experiments are limited to a narrow frequency
window~\cite{Wyss200x}. This will provide important insight into the
physical origins of the nonlinear viscoelastic response of a wide
range of soft materials.

The authors acknowledge support from the NSF (\# 0134969, KM and
DRR), (DMR-0602684, HMW and DAW), the Harvard MRSEC (DMR-0213805)
and the Swiss National Science Foundation (HMW).

\end{document}